\documentclass[sigconf]{acmart}

\usepackage{amsmath}
\usepackage{multirow}
\usepackage{tabularx}
\usepackage{graphicx} 
\usepackage{caption}
\usepackage{array}
\newcolumntype{C}[1]{>{\centering\arraybackslash}m{#1}}

\AtBeginDocument{%
  }

\setcopyright{acmlicensed}
\copyrightyear{2018}
\acmYear{2018}
\acmDOI{XXXXXXX.XXXXXXX}
\acmConference[Conference acronym 'XX]{Make sure to enter the correct
  conference title from your rights confirmation email}{June 03--05,
  2018}{Woodstock, NY}
\acmISBN{978-1-4503-XXXX-X/2018/06}

\copyrightyear{2026}
\acmYear{2026}
\setcopyright{cc}
\setcctype{by}
\acmConference[CHI '26]{Proceedings of the 2026 CHI Conference on Human Factors in Computing Systems}{April 13--17, 2026}{Barcelona, Spain}
\acmBooktitle{Proceedings of the 2026 CHI Conference on Human Factors in Computing Systems (CHI '26), April 13--17, 2026, Barcelona, Spain}
\acmPrice{}
\acmDOI{10.1145/3772318.3791105}
\acmISBN{979-8-4007-2278-3/2026/04}

\begin{document}

\title{How Much Trust is Enough? Towards Calibrating Trust in Technology}

\author{Gabriela Beltrão}
\orcid{1234-5678-9012}
\affiliation{%
  \department{School of Digital Technologies}
  \institution{School of Digital Technologies, Tallinn University}
  \city{Tallinn}
  \country{Estonia}}
  \email{gbeltrao@tlu.ee}

\author{Debora F. de Souza}
\affiliation{%
  \department{School of Digital Technologies}
  \institution{School of Digital Technologies, Tallinn University}
  \city{Tallinn}
  \country{Estonia}}
  \email{deboracs@tlu.ee}

\author{Sonia Sousa}
\affiliation{%
  \department{School of Digital Technologies}
  \institution{Tallinn University}
  \city{Tallinn}
  \country{Estonia}
}
\affiliation{%
  \department{Virumaa College}
  \institution{Tallinn University of Technology (TalTech)}
  \city{Kohtla-Järve}
  \country{Estonia}}
  \email{scs@tlu.ee}

\author{David Lamas}
\email{drl@tlu.ee}
\affiliation{%
  \department{School of Digital Technologies}
  \institution{Tallinn University}
  \city{Tallinn}
  \country{Estonia}}
  \email{drl@tlu.ee}

\renewcommand{\shortauthors}{Beltrão et al.}

\begin{abstract}
The role of trust within Human-Computer Interaction is being redefined. With the increasing omnipresence, autonomy, and opacity of technology, users often struggle to understand the capabilities and limitations of systems. In this article, we present the results of an empirical study designed to provide a practical, evidence-based interpretation of trust propensity assessment using the Human-Computer Trust Scale (HCTS). We outline the process used to develop a guideline for interpreting the instrument’s results and explain the rationale for our decisions, advocating for calibrating trust in technology within HCI. Our findings demonstrate that the HCTS is a promising tool for conducting an initial evaluation of propensity to trust, but that such an assessment requires reflection and interpretation that should be considered within the context of the interaction.
\end{abstract}



\maketitle

\section{Introduction}

Trust in technology is becoming increasingly important as our interactions with systems continue to evolve. Technology is no longer limited to specific moments of our days or confined to settings like sitting in front of a desktop computer. Instead, it has become integrated into our homes, cars, smartphones, and the streets we walk. Many of these systems operate autonomously, carrying out tasks without requiring supervision or direct input from individuals.

Following the evolution of technology, the role of trust within Human-Computer Interaction (HCI) is being redefined. Bodker’s notion of HCI waves \cite{bodker2006hciwaves} proposes that technology has progressively transformed from a resource primarily for specialized work to tools accessible to unspecialized users. This transition was accelerated by the emergence of Web 2.0 in the 2000s \cite{oreilly2005web20}, which transformed the role of technology from completing specific tasks to encompassing nearly all aspects of daily life \cite{bodker2015thirdwave}. More recently, advancements in artificial intelligence (AI) have further broadened and accelerated these changes. As a result, the nature of trust in technology has also evolved. 

While increasing trust was initially crucial for encouraging adoption and engagement with technology \cite{Fischer_2018}, the current landscape presents different challenges. In the face of technology's omnipresence, autonomy, and opacity, users often struggle to understand systems' capabilities and limitations. Consequently, instead of just addressing issues related to a lack of trust, HCI now faces further problems pertaining to overreliance and excessive trust. That has led researchers to underscore the importance of developing mechanisms to mitigate users' overtrust in autonomous technologies applied in sensitive contexts, such as healthcare \cite{Wang_2025OVERTRUST}. 

Considering the current challenges, our objective is to provide empirical evidence that supports more consistent assessment, interpretation, and reflective use of trust propensity measures. To achieve this, our work first examines trust propensity from a theoretical standpoint, demonstrating that, in today’s technology-driven environment, it is more important to understand the appropriate levels of trust rather than simply aiming for its highest levels.

Next, we present the results of two empirical studies that led to a practical, evidence-based guideline to complement the interpretation of the Human-Computer Trust Scale (HCTS), a validated tool for assessing the propensity to trust technology. We describe the process adopted for developing the guideline, the rationale behind our decisions, and suggestions on how to use it.

We base our work on the premise that trust is dynamic and responsive to the knowledge users gain about a system as interactions progress \cite{muir1987trust,Lewis_Weigert_1985,gaudiello2016trustrobot,miller2021, schoeller2021trust}. Thus, our goal is to encourage researchers and practitioners to examine users' propensity to trust at different stages and to reflect critically on the implications of their findings.

\section{Research Context}

\subsection{Trust Propensity and Trust Behavior}

Trust propensity refers to a person’s general tendency to trust, reflecting how their interpretation of a technology’s features shapes their willingness to rely on it. According to Mayer’s definition, trust propensity can be understood as the willingness to accept vulnerability based on expectations of another’s behavior \cite{mayer1995trust}. Alternatively, Lee and See \cite{lee2004trustautomation} conceptualize trust as a driver of behavior, viewing it as the belief that an agent will help achieve one’s goals amidst uncertainty and vulnerability. In this case, trust acts as an attitude that mediates interactions, shaping reliance and aiding users in navigating complex autonomous systems \cite{Hancock_2011, kok2020trust, lee2004trustautomation}. 

Previous research demonstrates that propensity to trust can influence actual behavior \cite{gefen2003trustTAM,gill2005antecedentstrust}, considering that trust develops through an attitude-formation process that can be adjusted by comparing expectations against the system’s actions and outcomes \cite{miller2021}. Nevertheless, the extent to which disposition translates into behavior remains uncertain, as it is influenced by various interrelated cognitive, behavioral, social, and cultural factors.

Despite this limitation, understanding propensity to trust remains crucial, as it helps foresee potential problems in the interaction \cite{kaplan2023trustmetaanaly} and prevent them from occurring. The present study focuses on this pre-interaction stage, which works as a baseline for subsequent interactions. We argue that prior attitudes need to be adequately interpreted to enable the recalibration of poorly adjusted trust dimensions \cite{muir1987trust, Lim_2023}. 

\subsection{From Fostering Trust to Calibrating It}
Improper use of technology is often underlaid by miscalibrated trust: excessive trust (overtrust) can result in misuse, while insufficient trust (distrust) may lead to disuse \cite{parasuraman1997humansautomation,kohn2021trustmeasurement, lee2004trustautomation}. Such imbalances lead to inefficient interactions or risks when users’ trust in technology exposes them to unnecessary dangers. Previous work \cite{visser2020trustcalibration} extends this understanding by highlighting that both overtrust and undertrust pose problems to the quality of the interactions, and that trust must be actively built, managed, and calibrated over time through strategies that can either lower excessive trust or address its insufficiency. Trust calibration aims to align users’ trust with a system’s actual capabilities. Achieving this requires maintaining an accurate and continuously updated perception of the system’s reliability and performance, refined through ongoing interaction and assessment \cite{muir1987trust,lee2004trustautomation,visser2020trustcalibration}. 

The notion of calibrated trust levels has been in focus of Ergonomics and Human Factors studies since the 1980s \cite{muir1987trust,leemoray1992trust,muir1996trustinatuomation, parasuraman1997humansautomation} for optimizing operators' performance during interactions with autonomous technology. Muir \cite{muir1987trust} argues that well-calibrated trust can be achieved by enhancing the perception of trustworthiness, offering explicit meanings for users to adjust the trustworthiness criteria, supporting users to allocate functions and, respectively, their reliance on the systems, and, lastly, recalibrating poorly adjusted trust dimensions (designing, evaluating, and iterating). Hence, calibration requires understanding both trust as a predisposition, "specifying the expectation" (\cite{muir1987trust}, p. 537), and as a continuously updating attitude which guides the improvement of interaction.

Still, to calibrate trust, it is necessary to measure it. Yet, a fundamental challenge remains: assessing trust is difficult because most measurement tools are context-dependent \cite{Tenhundfeld_2022, kohn2021trustmeasurement}. In the current technological landscape, marked by increasing complexity, a lack of transparency, and a rapid pace of adoption driven by external pressures, our research advocates for a trust calibration approach, shifting away from unidirectional efforts to increase trust in interactions.

\subsection{Measuring Propensity to Trust Technology with the Human-Computer Trust Scale (HCTS)}

The HCTS is a scale developed for measuring propensity to trust in technology \cite{gulati2019hctm}, based on the conceptual Human-Computer Trust Model (HCTM) \cite{sousa2014trustmodel, gulati2017modelling}. The model considers trust propensity from a socio-technical perspective \cite{emery1960socio}, that is, examining it as part of social and organizational structures. The model and subsequent tool are based on Mayer's \cite{mayer1995trust} conceptualization of trust, which is grounded in individuals' subjective assessments of the risks and benefits involved in an interaction.

The HCTS consists of a psychometric questionnaire designed to measure trust propensity by assessing respondents' agreement with a series of statements on a 5-point Likert scale. These statements pertain to different trust constructs, and their general mean yields the trust score. However, the interpretation of the results is primarily context-specific. It is possible to compare the results of different trust constructs to understand how they influence the trust levels, and to compare results among groups of respondents within the study to investigate their differences in trust propensity.

While the single score produced by the HCTS is useful for these relative comparisons, interpreting the results in more general terms remains a challenge. First, it is problematic to compare scores across different assessments because trust propensity is context-dependent, and the study design, particularly the stimuli used, can influence individuals’ perceptions. Second, although the instrument has been applied across various contexts of technology use \cite{gulati2019hctm,oper2020trust,fimberg2020trust,pinto2020hctm, pinto2022trust,beltrao2021trustwhatsapp}, these studies have been analyzed in isolation, without efforts to establish more general grounds for interpretation, limiting the applicability of the scale. 

Our goal is to fill this gap and contribute to the development of a tool that can be more widely used. We focus on the HCTS because it is a theoretically grounded instrument that has been validated and refined, and because its replication across diverse contexts indicates its suitability as a potential general tool for assessing individuals’ propensity to trust technology. To address challenges in interpreting results, we propose establishing thresholds that complement the existing process and encourage researchers to reflect on their results. 

In the following sections, we will outline the process we used to identify general ranges that can aid in the interpretation of HCTS results. We adopt the revised questionnaire \cite{beltrao2025micom}, which includes nine items measuring four constructs: 

\begin{itemize}
    \item \textbf{Competence (COM):} the perception of a system's ability to perform its intended tasks effectively by providing the appropriate features and functionalities, thereby meeting user expectations.
    
    \item \textbf{Benevolence (BEN):} the perception that the technology acts in the user’s best interest, even when there is no obligation or reward for doing so \cite{bhattacherjee2002trustonline}. In practical terms, it means that the system is designed to provide adequate support to help users achieve their interaction goals \cite{gulati2019hctm}. 

    \item \textbf{Perceived Risk (PR):} individuals’ subjective evaluation of the potential consequences that may arise from negative experiences while interacting with technology. This concept is rooted in the work of Pavlou and Gefen \cite{pavlougefen2004buildingtrust}, who define Perceived Risk as the subjective belief in the possibility of loss due to potential opportunistic behaviors by (socio-technical) systems.
    
    \item \textbf{Structural Assurance (SA):} the belief that reliable legal, contractual, or physical mechanisms are in place to support and secure the use of technology. It is related to institution-based trust, thus a disposition related to the broader structure in which the technology is used \cite{mcknight2002developingtrust}.  
\end{itemize}

\section{Methodology}

Our approach is inspired by the process adopted for the development of a range for the interpretation of the System Usability Scale (SUS). The SUS is a widely used instrument in usability research, which serves as a simple yet reliable tool for measuring systems' usability \cite{brooke1996sus}. It uses a 10-item Likert-scale questionnaire, with results ranging from 0 to 100.

After more than a decade of usage of the SUS, efforts were put forth towards translating the numeric results (from 0 to 100) into an acceptability range ("not acceptable", "marginal", or "acceptable"), aiming to facilitate interpretation  \cite{bangor2008empirical}. To reach a reliable acceptability rate, the authors analyzed data from SUS surveys in which participants rated usability using both the SUS and a 7-point adjective scale (e.g., "Poor," "Good," "Excellent"). The acceptability rate is the outcome of the authors' mapping of the score ranges to descriptive terms based on their correlation and alignment with the conceptual understanding of what the scale measures \cite{bangor2009determining}.

We draw inspiration from this procedure to develop a range that helps practitioners interpret HCTS scores. Still, we recognize that it does not entirely overcome the subjectivity of the evaluation. Just as the SUS authors state that usability does not exist in an absolute sense, but rather refers to "appropriateness to a purpose" \cite[p. 2]{brooke1996sus}, we emphasize that the propensity to trust must be interpreted in context and in relation to the system and its functions.

\subsection{Creation of the Adjective Scale}

The first step in our procedure was to include an adjective scale in the studies conducted with the HCTS. This scale required participants to rate how much they trusted the technology investigated using a set of adjective descriptions, after the HCTS statements.

The adjective description followed the criteria proposed by Dodd \& Gerbrick \cite{dodd1960word}, namely by aiming for phrases: (1) at a unidimensional scale, (2) at equal intervals, (3) which has the least ambiguity or dispersion, (4) of greatest familiarity to most people, and (5) which are logically symmetric.

Although it is problematic to claim that adjectives are at "equal intervals", we relied on existing rating scales to guide the selection of phrases that represent the degrees of agreement or disagreement based on the relative degree of "goodness" of the terms \cite{babbitt1989questionnaire}. We selected the adjectives based on studies on descriptors to include in questionnaires \cite{matthews1978perceived}, using seven alternatives for greater granularity. The final scale adopted in the studies is presented in Table \ref{tab:adjectives}:

\begin{table}[!htp]\centering
\caption{Adjective scale used in the studies}\label{tab:adjectives}
\begin{tabular}{lc}\toprule
\textbf{How would you describe your trust in [system]?} \\\midrule
I trust it completely \\
I mostly trust it \\
I trust it moderately \\
I am not sure if I trust it \\
I somewhat do not trust it \\
I mostly do not trust it \\
I completely do not trust it \\
\bottomrule
\end{tabular}
\Description{Respondents were asked, “How would you describe your trust in [system]?” and provided with seven ordered response options ranging from complete trust to complete distrust. The scale included the following categories: I trust it completely, I mostly trust it, I trust it moderately, I am not sure if I trust it, I somewhat do not trust it, I mostly do not trust it, and I completely do not trust it.}
\end{table}

The goal of the adjective scale is to help us identify cut-offs that can aid the interpretation of the numeric results. Nevertheless, guided by the idea of calibration \cite{muir1987trust, Tenhundfeld_2022, visser2020trustcalibration}, we argue that, as a general notion, propensity to trust should not be the highest possible, but be in a moderate-to-high range (I trust it moderately, I mostly trust it). Therefore, the low end of the scale (I completely do not trust it, I mostly do not trust it) represents undertrust and indicates that there might be problems with the system. In contrast, the highest end of the scale (I trust it completely) suggests that users may overtrust the system, which also points to an issue that might require attention.

\subsection{Data Collection}

This article relies on the aggregated data collected in a series of online surveys conducted for two independent studies between 2023 and 2025, investigating propensity to trust in two systems: (1) Facial Recognition Systems for Law Enforcement and (2) Biometric Payment Systems. The studies had differences in some of the items, but were identical in the content regarding this work. Each study included a stimulus and relied on the participants' perceptions of the hypothetical implementation of the systems. In both studies, the stimulus was a short video presenting the technology, with 2:10 and 3:36 minutes, respectively. The surveys were conducted online using the platform LimeSurvey (http://www.limesurvey.org/), and the results were later aggregated. The participants were recruited through convenience sampling.

\begin{itemize}
    \item \textbf{Study 1: Facial Recognition Systems for Law Enforcement}: The stimulus consisted of two video excerpts, one about Skynet (China) and the other about the Metropolitan Police of London (England). Each excerpt illustrates real uses with slightly different approaches. Skynet emphasizes the system's efficiency, while the Metropolitan Police highlights how the citizens' privacy is maintained. The study comprises data aggregated from 7 surveys.
    \item \textbf{Study 2: Biometric Payment Systems}: The stimulus consisted of a short news segment about Biometric Payments using facial recognition, explaining the technology's functioning, risks, and benefits. The study includes data aggregated from 5 surveys.
\end{itemize}

Despite the technical similarity between the systems adopted in both studies, they differ in their purpose, potential risks, and benefits to the users, making them relevant cases for comparison. In the surveys, trust propensity was assessed using the revised HCTS \cite{beltrao2025micom}, with a total of 9 items measured on a 5-point Likert scale, comprising the four constructs previously described. Additionally, the participants were asked to rank how they would describe their trust as per the adjective scale (Table \ref{tab:adjectives}).

\subsection{Data Analysis}
To identify adjective categories with significant differences, we employed an iterative process with one-way ANOVAs to determine if the categories differed, followed by post-hoc tests to assess whether all groups were significantly different from each other, in each study. If the adjective categories were significantly different in both studies, they were maintained. If not, we examined which groups did not satisfy the criteria and, if conceptually plausible, merged the categories (Steps 1 and 2). Next, in Step 3, we calculated cut-offs across studies. The steps are detailed next.

\subsubsection{Step 1: Within-Study Label Estimation}
First, we established cut-offs for the HCTS scores corresponding to the seven adjective categories. This categorization was performed separately for the two studies included in the analysis, as each dataset exhibited distinct distributions and sample characteristics. The mean, standard deviation (SD), and sample size (N) for each study were used to calculate the standard error (SE). Next, we calculated the 95\% confidence intervals (CIs) around each group’s mean score \cite{wasserman2013statistics}: 

\[CI_{95\%}= Mean \pm 1.96 \times SE\] 

\subsubsection{Step 2: Within-Study Boundaries Estimation}
To estimate the boundary between two adjacent adjective categories (e.g., I completely do not trust it, I somewhat do not trust it), we calculated the midpoint between the upper bound of the lower category's CI and the lower bound of the upper category's CI:

\[Cut-off = \frac{(UpperCI_i + LowerCI_{i+1})}{2}\]

This approach aimed to avoid arbitrary divisions and ensure that the boundaries reflect statistically distinguishable ranges, as long as supported by prior ANOVA analyses confirming that the differences are statistically significant.

\subsubsection{Step 3: Defining General Cut-offs Across Studies}

After deriving cutoffs separately for each study, we proposed a general set of cutoffs to be applied across both studies. We relied on the combination of the studies’ results, but recognize that there are limits to generalization due to the highly contextual nature of trust and the inherent subjectivity of the interpretation. 

\section{Results}
We present the results from the two studies, Study 1 (N = 711), and Study 2 (N = 227). Cronbach's Alphas for the HCTS items were 0.814 and 0.860 for Studies 1 and 2, respectively. Prior analyses confirmed the HCTS partial measurement invariance across countries, supporting the instrument's stability for relational analyses \cite{beltrao2025micom}. The overview of the dataset for each study is presented in Table \ref{tab:descriptives}.

\begin{table}[h!]\centering
\caption{Sample characteristics for Studies 1 and 2}\label{tab:descriptives}
\begin{tabular}{C{1.7cm}C{2.1cm}cc|cc}\toprule
& &\multicolumn{2}{c|}{\textbf{Study 1}} &\multicolumn{2}{|c}{\textbf{Study 2}} \\\cmidrule{3-6}
& &N &\% &N &\% \\\midrule
\textbf{Total} & &\textbf{711} &- &\textbf{227} &- \\\midrule
\multirow{3}{*}{Gender} &Male &325 &45.7 &101 &44.5 \\
&Female &378 &53.2 &124 &54.6 \\
&Other &8 &1.1 &2 &0.9 \\\midrule
\multirow{5}{*}{Age group} &24 or less &281 &39.5 &39 &17.2 \\
&25-34 &172 &24.2 &117 &51.5 \\
&35-44 &161 &22.6 &58 &25.6 \\
&45-54 &70 &9.8 &11 &4.8 \\
&55 or more &27 &3.8 &2 &0.9 \\\midrule
\multirow{3}{*}{Education} &Basic &92 &12.9 &7 &3.1 \\
&Higher &457 &64.3 &166 &73.1 \\
&Postgraduate &162 &22.8 &54 &23.8 \\\midrule
\multirow{5}{*}{Region} &Africa &0 &0 &33 &14.5 \\
&Asia &454 &63.9 &57 &25.1 \\
&Europe &117 &16.5 &91 &40.1 \\
&South \& Central America &133 &18.7 &27 &11.9 \\
&North America &7 &1 &19 &8.4 \\
\bottomrule
\end{tabular}
\Description{Table summarizing sample characteristics for Study 1 (N=711) and Study 2 (N=227). For both studies, the table reports distributions by gender, age group, education level, and region. In both samples, female respondents slightly outnumber male respondents, and the majority of participants are under 35 years old. Most respondents have higher or postgraduate education. Study 1 is dominated by participants from Asia, whereas Study 2 includes a more regionally diverse sample with substantial representation from Europe, Asia, and Africa.}
\end{table}

\subsection{Differences in Scores}
We conducted two iterations to reach statistically significant differences in the adjective categories in all cases, in both studies. Study 1 showed statistically significant differences in scores between the groups from the first iteration (7 categories). Study 2, on the other hand, did not meet the criteria for four pairs in the original scale and for three pairs in the first iteration. In the second (final) iteration, results were significantly different for all pairwise comparisons for both studies.

\subsubsection{Adjective Scale Recoding}
\begin{figure}[h]
    \centering
    \includegraphics[width=\linewidth]{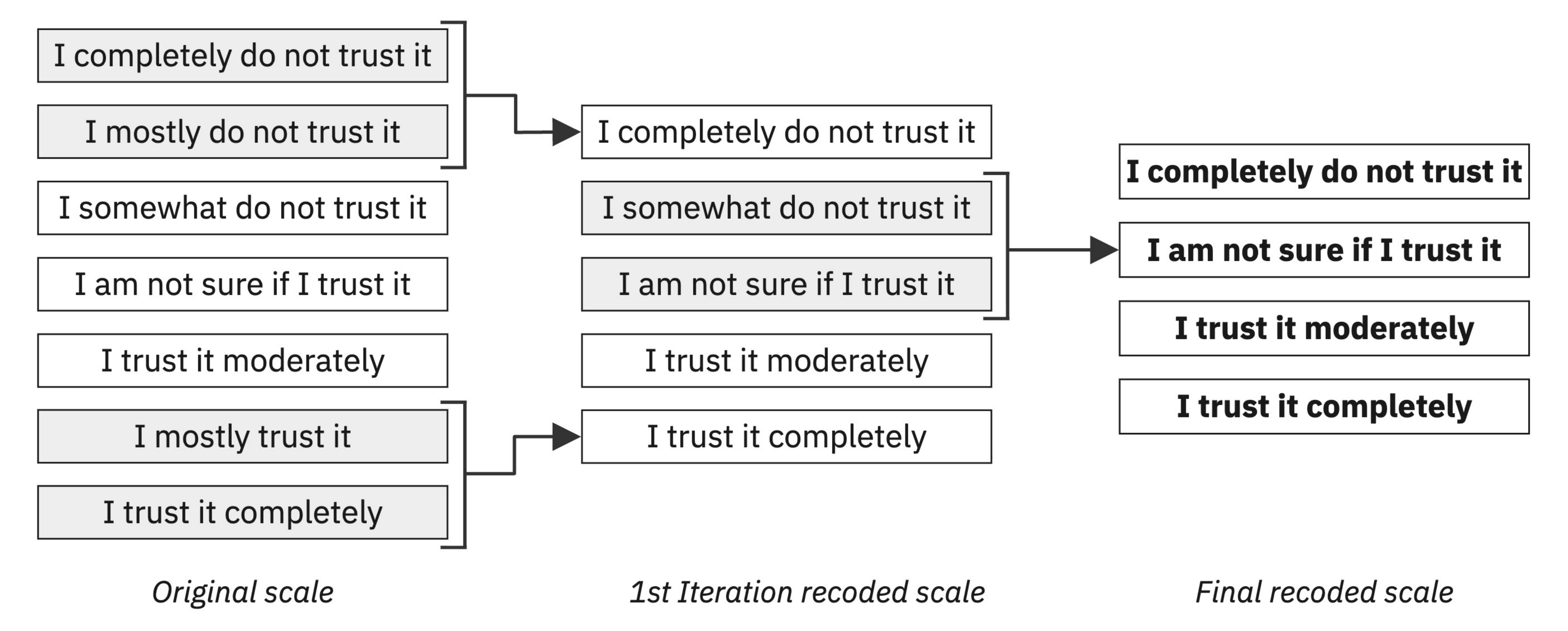}
    \caption{Summary of recoding procedure}
    \label{fig:recoding1}
    \Description {Diagram illustrating the recoding procedure of the trust scale across three stages. The figure shows the original seven-point trust scale, an intermediate recoded version, and a final four-point scale, highlighting how adjacent trust categories are progressively merged to increase interpretability. The final recoded scale consists of four ordered categories ranging from complete distrust to complete trust.}
\end{figure}

The process for recoding the scale is presented in Figure \ref{fig:recoding1}, where the original adjective scale, used in the data collection, is presented on the left, and the final recoded one on the right. In all cases, a lack of statistically significant differences occurred only between adjacent groups. For clarity, we present the ANOVA results only for the final iteration (final recoded scale). The dataset and data analysis files are available from the corresponding author upon request.

For the final recoded scale, the one-way ANOVA revealed significant differences between the categories for both studies, F(3, 695) = 46.745, p < .001 and F(3, 213) = 13.273, p < .001, Levene’s test indicated a violation of the homogeneity of variances assumption in both cases (p < .001), so pairwise comparisons were conducted using the Games-Howell post-hoc test. With four groups considered statistically different, we proceeded to define the cut-offs for the intervals. 

\subsubsection{Defining Boundaries}
To identify score intervals corresponding to four adjective categories, we calculated the mean and 95\% confidence intervals for the categories in each study. The cut-offs were placed at the midpoints between adjacent category means, ensuring alignment with the differences between the categories. Tables \ref{tab:cutoffs1} and \ref{tab:cutoffs2} present the complete cut-off results.

\begin{table*}[!htp]  
\centering
\caption{Summary of cut-offs calculation for Study 1}\label{tab:cutoffs1}

\begin{minipage}{\linewidth}
\centering
\begin{tabular}{lccccccccc}\toprule
& & & & & &\multicolumn{2}{c}{\textbf{95\% CI}} & \\\cmidrule{7-8}
\textbf{Final recoded scale} &\textbf{Mean} &\textbf{N} &\textbf{SE} &\textbf{WM} &\textbf{SE(WM)} &\textbf{LB} &\textbf{UB} &\textbf{Cut-off} \\\midrule
\textbf{I completely do not trust it} &1.98 &50 &0.09 &1.98 &0.07 &1.85 &2.11 &\textbf{2.43} \\
\textbf{I am not sure if I trust it} &2.79 &187 &0.04 &2.79 &0.02 &2.75 &2.84 &\textbf{2.96} \\
\textbf{I trust it moderately} &3.12 &281 &0.03 &3.12 &0.02 &3.09 &3.15 &\textbf{3.39} \\
\textbf{I trust it completely} &3.68 &181 &0.04 &3.68 &0.02 &3.63 &3.72 &\textbf{} \\
\bottomrule
\end{tabular}

\vspace{2pt}
    {\footnotesize\centering
      SE = Standard Error, WM = Weighted Mean, LB = Lower Bound, UB = Upper Bound
      \par
    }
\Description{Table summarizing how cut-off values were calculated for the recoded trust scale in Study 1, reporting mean scores, sample sizes, standard errors, weighted means, and 95 percent confidence intervals for each category, and indicating the derived thresholds that separate adjacent levels of trust from complete distrust to complete trust.}
\end{minipage}
\end{table*}

\begin{table*}[!htp]
 
\centering
\caption{Summary of cut-offs calculation for Study 2}\label{tab:cutoffs2}

\begin{minipage}{\linewidth}
\centering
\begin{tabular}{lccccccccc}\toprule
& & & & & &\multicolumn{2}{c}{\textbf{95\% CI}} & \\\cmidrule{7-8}
\textbf{Final recoded scale} &\textbf{Mean} &\textbf{N} &\textbf{SE} &\textbf{WM} &\textbf{SE(WM)} &\textbf{LB} &\textbf{UB} &\textbf{Cut-off} \\\midrule
\textbf{I completely do not trust it} &2.28 &21 &0.14 &2.28 &0.09 &2.10 &2.45 &\textbf{2.60} \\
\textbf{I am not sure if I trust it} &2.81 &61 &0.06 &2.81 &0.04 &2.74 &2.88 &\textbf{2.96} \\
\textbf{I trust it moderately} &3.11 &73 &0.07 &3.11 &0.04 &3.04 &3.18 &\textbf{3.37} \\
\textbf{I trust it completely} &3.67 &62 &0.10 &3.67 &0.05 &3.57 &3.78 &\textbf{} \\
\bottomrule
\end{tabular}

\vspace{2pt}
    {\footnotesize\centering
      SE = Standard Error, WM = Weighted Mean, LB = Lower Bound, UB = Upper Bound
      \par
    }

\Description{Table summarizing the calculation of cut-off values for the recoded trust scale in Study 2. For each trust category, the table reports mean scores, sample sizes, standard errors, weighted means, and 95 percent confidence intervals, and shows the derived cut-off thresholds that separate adjacent levels of trust from complete distrust to complete trust.}
\end{minipage}
\end{table*}

\subsection{Interpretation of HCTS Results}
The derived cut-offs across the two studies showed small variability, upholding the robustness of these boundaries. To support the practical application and understanding of HCTS results, we calculated cut-offs based on the CI estimates from the average results of both studies, which are presented in Table \ref{tab:generalbounds}. 

\begin{table}[!htp]\centering
\begin{minipage}{\linewidth}
\centering
\caption{Bounds and cut-off based on Studies 1 and 2}\label{tab:generalbounds}
\begin{tabular}{lcccc}\toprule
&\multicolumn{2}{c}{\textbf{95\% CI}} & \\\cmidrule{2-3}
&\textbf{LB} &\textbf{UB} &\textbf{Cut-off} \\\midrule
\textbf{I completely do not trust it} &1.97 &2.28$^{*}$ &2.51 \\
\textbf{I am not sure if I trust it} &2.74 &2.86 &2.96 \\
\textbf{I trust it moderately} &3.06 &3.17 &3.38 \\
\textbf{I trust it completely} &3.60$^{*}$ &3.75 & \\
\bottomrule
\end{tabular}

\vspace{2pt}
    {\footnotesize\centering
      LB = Lower Bound, UB = Upper Bound \par
       $^{*}$ Values used for the interpretation ranges
      \par
    }

\Description{Table synthesizing bounds and cut-off values for the recoded trust scale based on results from Studies 1 and 2. For each trust category, the table reports the lower and upper bounds of the 95 percent confidence interval and the final cut-off used for interpretation. Starred values indicate the selected thresholds that define interpretation ranges across the two studies, progressing from complete distrust to complete trust.}

\end{minipage}
\end{table}

\begin{figure}[h!]
    \centering
    \includegraphics[width=\linewidth]{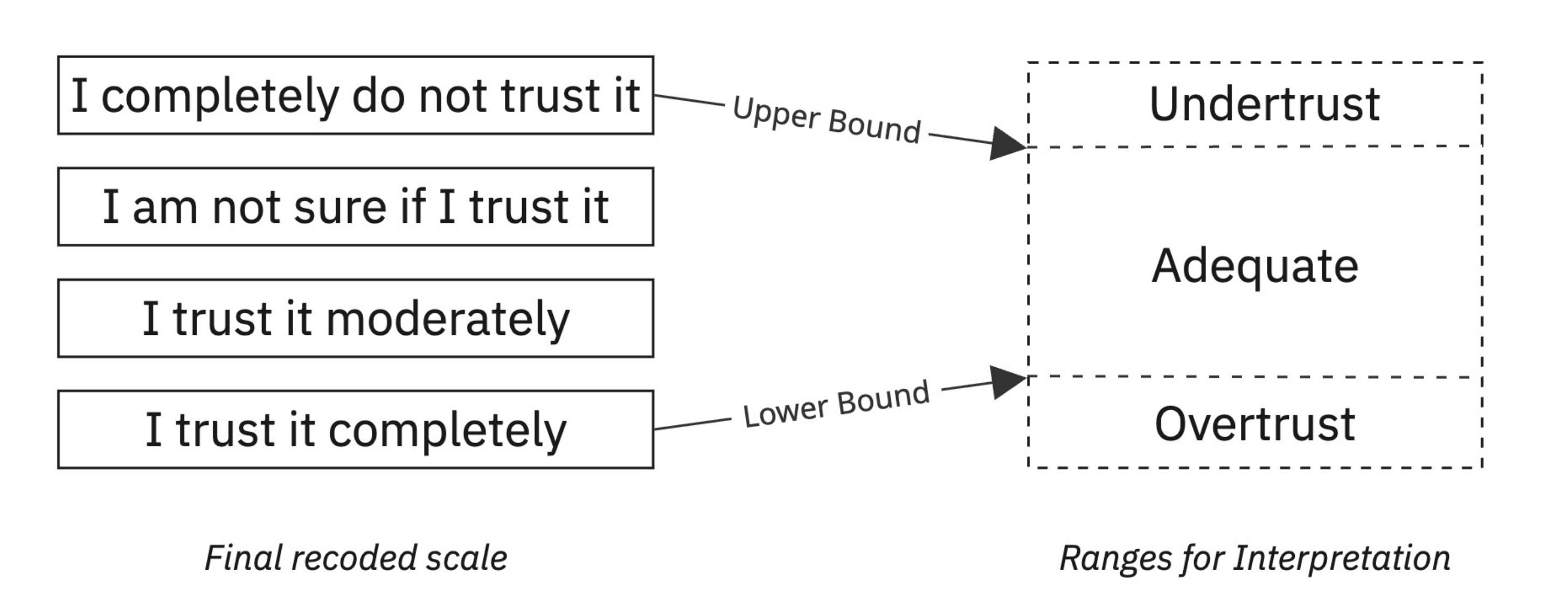}
    \caption{Overview of the procedure adopted for defining bounds}
    \label{fig:procedurerange}
    \Description{Diagram showing how lower and upper bounds are used to map the final recoded trust scale onto three interpretive categories: undertrust, adequate trust, and overtrust. The figure illustrates how the ordered response options correspond to these ranges across the scale.}
\end{figure}

In line with our conceptual framing \cite{muir1987trust, lee2004trustautomation, visser2020trustcalibration}, we propose that a moderate level of trust propensity, rather than the highest level, is more appropriate in most cases. Thus, we adapted the results in two steps. First, we translated the cut-offs based on individuals’ self-reported trust into ranges within the HCTS. Figure \ref{fig:procedurerange} illustrates the process employed for transitioning from the ranking to the HCTS interpretation.

Next, to facilitate implementation and ensure flexibility, we used the upper and lower bounds of the 95\% CIs surrounding the estimated category transition points to define the score ranges for each category. We rounded the values to facilitate the interpretation and practical usage of the ranges. This approach was adopted to ensure that the categories are empirically grounded and distinguishable, but also reflect the inherent variability of what can be considered the adequate propensity to trust. 

Table \ref{tab:interpretation} presents the overview of the values, based on the CIs as shown in Table \ref{tab:generalbounds} and the procedure explained in Figure \ref{fig:procedurerange}. Additionally, we propose Figure \ref{fig:hcts-colors} as a reference for interpreting the results, which more adequately depicts the score as falling on a continuum with transition areas rather than separated by fixed boundaries. A quick guide on how to use the HCTS and interpret its results is available in Appendix \ref{app:how-to-hcts}.

\begin{table}[!htp]\centering
\caption{Proposed HCTS interpretation.}\label{tab:interpretation}
\begin{tabular}{lcc}\toprule
\textbf{Interpretation Ranges} &\textbf{Value} \\\midrule
Undertrust & $\leq$ 2.30 \\
Adequate &2.31-3.60 \\
Overtrust &$\geq$ 3.61 \\
\bottomrule
\end{tabular}
\Description{Table defining interpretation ranges for the Human–Computer Trust Scale. The table groups trust scores into three interpretive categories (undertrust, adequate trust, and overtrust) based on the final cut-off values, providing a simplified framework for interpreting trust scores in subsequent analyses.}
\end{table}

\section{Discussion}

The results of our empirical study point to key considerations for the assessment and interpretation of the HCTS. First, there was a lack of variance between adjacent groups in the initial iterations of the interpretation categories, indicating that while the thresholds capture meaningful distinctions, these differences might not be categorical. The fact that the differences were not observed in Study 2, which has a smaller sample, is also a result of the methods' sensitivity to sample size \cite{kraemer2015statanalysis}. However, although a smaller sample is often viewed as a limitation due to reduced statistical power and generalizability, it allowed us to identify group-specific differences that might be harder to discern in larger samples, where individual or subgroup variations can be obscured by aggregation, and small effects may become statistically significant. Therefore, we view Study 2 as an important complement that offers exploratory insight into the differences across groups.

\begin{figure}[htb!]
    \centering
    \includegraphics[width=\linewidth]{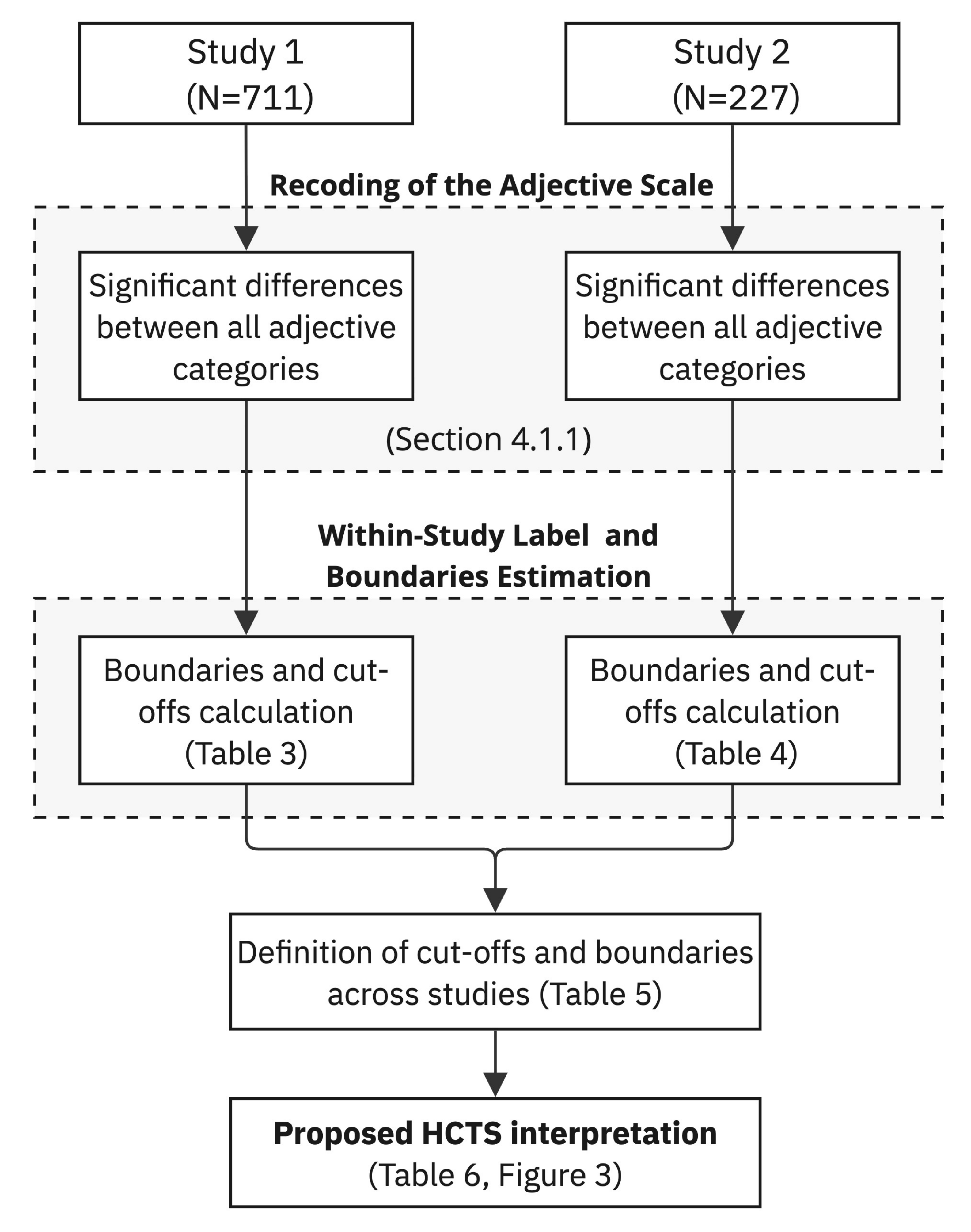}
    \caption{Overview of the workflow across studies, from adjective scale recoding to the proposed HCTS interpretation}
    \label{fig:procedureoverview}
    \Description{Flowchart of the analytical pipeline for two studies, showing parallel processes for Study 1 (N=711) and Study 2 (N=227) that begin with recoding the adjective scale and identifying significant differences between categories, proceed to within-study label and boundary estimation with cut-offs calculation, and conclude with a unified definition of cut-offs and a proposed interpretation of the HCTS.}
\end{figure}

While some distinctions between categories did not achieve statistical significance across all datasets initially, the final structure allows for a nuanced interpretation, which aligns with conceptually meaningful distinctions that remain consistent across studies. Figure \ref{fig:procedureoverview} presents an overview of the procedure adopted to reach the proposed interpretation, indicating where the results were integrated.  

\begin{figure*}[h]
    \centering
    \includegraphics[width=.95\linewidth]{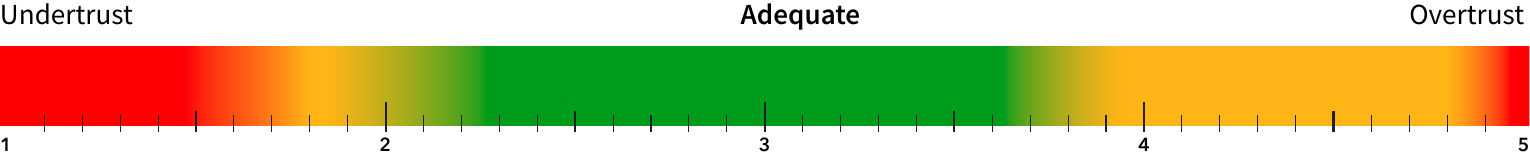}
    \caption{Visual representation of proposed interpretation range}
    \label{fig:hcts-colors}
    \Description{Horizontal visual scale illustrating the proposed interpretation ranges for the Human–Computer Trust Scale. The figure shows a continuous progression across three interpretive categories (undertrust, adequate trust, and overtrust) using visual transitions in colors to emphasize that category boundaries are gradual and intended as interpretive guides rather than strict thresholds.}
\end{figure*}

Next, it is noteworthy that the observed range of values tends to center around three, which is the midpoint of the scale. This trend raises questions about whether adequate trust leans toward neutrality, or if this pattern may have emerged from the combination of responses needed to establish statistical differences between categories (as shown in Figure \ref{fig:recoding1}). In this case, the trend may have been influenced by our conceptual framing of what constitutes adequate trust, which considers both ends of the scale as needing attention--although undertrust tends to be the more pressing concern. 

These findings underscore a need for a more nuanced understanding of how trust manifests across different contexts and whether neutrality truly reflects a moderate propensity to trust or strategic disengagement. Therefore, more efforts are needed to reach both a more in-depth understanding of the factors that affect individuals’ propensity to trust and of the mechanisms by which it can be increased or dampened. Still, our framing aimed to prioritize the practicality of the tool and the flexibility of the interpretation. 

Similarly, adopting the 95\% CI bounds, instead of the cutoffs, broadened the “Adequate” trust range to better accommodate the contextual nature of trust propensity \cite{Hoff_2014, miller2021,beltrao2023trustfrs}. As such, the range, as presented in Figure \ref{fig:hcts-colors}, establishes a baseline for the interpretation of the HCTS results while, at the same time, encourages reflection on what constitutes adequate trust in a given situation.

The fact that the extreme adjectives had significant differences further suggests that they reflect distinct mindsets, making them suitable points of reference for interpreting these tendencies. This decision follows the general understanding that both too little and too much trust can be problematic. Our rationale follows the calibration model by Lee and See \cite{lee2004trustautomation}, grounded in the concept of resolution, which informed our conceptualization of the trust boundaries. In this view, good resolution appears when a user’s level of trust aligns with the system’s actual capabilities, supporting appropriate reliance. On the other hand, over- or undertrust describes poor resolution, where the user’s trust appraisal either exceeds or falls short of what the system can actually deliver. We therefore treated the lowest and highest adjective ratings as indicative anchors of under- and overtrust, rather than as direct behavioral markers. 

The interpretation range is not normative and does not replace a detailed analysis of the assessment; it guides an initial interpretation that should be complemented by considering the trust constructs and further comparative analyses. We emphasize that researchers should be cautious about overinterpreting the results, especially considering the sensitivity to context and sample size. Even so, the proposed threshold marks a significant improvement in the interpretation of trust propensity with the HCTS. We foresee the assessment mechanism of our study as a valuable instrument for both research and educational purposes.

\section{Limitations and Future Studies}
Our findings should be considered in light of some limitations. First, the analyses were limited to two studies, focusing on two technologies that are similar from a technical standpoint, which are far from depicting the heterogeneity of existing technological systems and interaction contexts. In addition, invariance was not assessed across contexts due to differences in sample sizes, limiting comparability. The presented findings should therefore be interpreted with caution and not overgeneralized.

Second, the procedure for developing thresholds inherently involves a high degree of subjectivity. Despite our efforts to follow the conceptual frameworks presented, further research is needed to reify these thresholds across a broader range of domains. Building on this point, future research should investigate whether the observed “adequate” range, truly reflects appropriate levels of trust or if these values need to be refined. 

More studies are also required to explore in more depth the interplay between the trust constructs and adequate trust. Furthermore, given the nature of the HCTS assessment, which allows for a breakdown of the trust constructs, future studies could further analyze them separately to investigate how each dimension affects the overall score. In sum, our research prioritized an exploratory approach to advance towards a practical tool, and we acknowledge that additional investigations, both in breadth and depth, are necessary to enhance the robustness of our findings.

\section{Conclusion}

Trust in technology has taken center stage in discussions about the risks of emergent autonomous technologies. In the face of such high interest, HCI researchers have struggled to find easy-to-use research instruments that do not inherit replicability issues (e.g., often narrow or contextually bound questions) \cite{desouza2025HRItrust, Hancock_2011,miller2021,bach2022reviewtrustai}. In this article, we discussed the evolution of trust in technology in HCI and used empirical evidence to provide means for advancing towards a tool for measuring and interpreting individuals' trust propensity better aligned with current needs.   

Drawing on data from two studies, we propose guidelines for assessing and interpreting trust propensity to assist researchers and practitioners in analyzing it. For that, we take inspiration from the concept of trust calibration \cite{muir1987trust, lee2004trustautomation} and from the procedure used to develop interpretation ranges for the SUS \cite{bangor2009determining}. 

Addressing these challenges, we presented the HCTS as a promising and accessible tool for conducting an initial assessment of trust propensity, proposing an interpretive range that explores this tool's potential for practice and critical reflection.

\begin{acks}
We acknowledge the support of Estonian Research Council, project TEM-TA26 “Digital Transformation Through Life-Long Learning”, funding program Temaatilised teadus- ja arendusprogrammid (TEM-TA); and Education and Youth Board, project ÕÜF9 "Development of robot-human co-creation in industry”, funding program Õiglase Ülemineku Fondi meede (ÕÜF).
\end{acks}

\bibliographystyle{ACM-Reference-Format}
\bibliography{bibliography1}

\appendix

\section{How to use the HCTS and interpret the results}
\label{app:how-to-hcts}

The Human Computer Trust scale (HCTS) is a quick-and-dirty psychometric scale that assesses individuals' predisposition to trust a technological artifact. This instrument can be used for exploratory and descriptive evaluation, focusing on a given context at a moment in time.
To facilitate its use by researchers, designers, and stakeholders, we present our scale utilizing placeholders "[------]". These placeholders should be replaced with the specific system being evaluated. The additional placeholders should be replaced in COM1 and COM2 with the technology's function, and in SA2 with the type of technology.

The scale consists of 9 items referring to 4 constructs: Perceived Risk (PR), Benevolence (BEN), Competence (COM), and Structural Assurance (SA).  Responses should be measured on a 5-point Likert scale with anchors: "1-Strongly disagree" and "5-Strongly agree". PR is a negative measure, so its items (marked with "*") should have their results reversed. The items are as follows:

\begin{itemize}
    \item \textbf{COM1}: [------] are competent and effective in [------].
    \item \textbf{COM2}: [------] performs its role in [------] very well.
    \item \textbf{PR1}*: There could be negative consequences when using [------].
    \item \textbf{PR2}*: It is risky to interact with [------].
    \item \textbf{BEN1}: [------] will act in my best interest.
    \item \textbf{BEN2}: [------] will do its best to help me if I need help.
    \item \textbf{BEN3}: [------] are interested in understanding my needs and preferences.
    \item \textbf{SA1}: I feel assured that legal and technological structures provided by the government protect me when using [------].
    \item \textbf{SA2}: I can trust [------] because [------] regulations are robust and safe.
\end{itemize}

The interpretation of results is primarily exploratory, aiming to predict users' predisposition to trust the chosen technology. To calculate the trust scores, first reverse the PR (marked with *) scores by subtracting each score from 6. For example, a response of 4 would be reversed to 2. For all other items, accept the responses as they are. Next, calculate the mean for all questions; this mean represents the \textbf{trust score}, which can be interpreted using the ranges provided in Figure \ref{fig:hcts-colors} as a reference. This is an initial range for reference, which must be complemented with a more in-depth interpretation of the results.

Next, researchers should calculate the mean for each construct (RP, COM, BEN, SA) using their respective items. It is also possible to compare scores across different groups (e.g., demographic, behavioral) to explore variations in trust propensity among respondents. In this case, the \textbf{trust score} can be considered a baseline.

\end{document}